\documentclass[aps,
 preprint,
 preprintnumbers,
 preprintnumbers,
 nobibnotes,
 bibnotes,
 amsmath,amssymb,
 aps,
 floatfix,
]{revtex4-1}

\usepackage{graphicx}% Include figure files
\usepackage{dcolumn}% Align table columns on decimal point
\usepackage{bm}% bold math
\usepackage{natbib}
\usepackage{xcolor}
\usepackage{ulem} 
\usepackage{gensymb}
\usepackage{appendix}

\begin{document}

\title{Pressure-induced metal-insulator transition in oxygen-deficient LiNbO$_3$-type ferroelectrics }% Force line breaks with \\

\author{Chengliang Xia$^{1,2}$, Yue Chen$^{1}$ and Hanghui Chen$^{2,3}$}
%\email{Correspondence to hanghui.chen@nyu.edu or yuechen@hku.hk.}
\affiliation{
  $^1$Department of Mechanical Engineering, The University of Hong Kong, Pokfulam Road, Hong Kong SAR, China \\
  $^2$NYU-ECNU Institute of Physics, NYU Shanghai, Shanghai, 200062, China\\
  $^3$Department of Physics, New York University, New York  10003, USA\\
}


\begin{abstract}
Hydrostatic pressure and oxygen vacancies usually have deleterious
effects on ferroelectric materials because both tend to reduce their
polarization. In this work we use first-principles calculations to
study an important class of ferroelectric materials -- LiNbO$_3$-type
ferroelectrics (LiNbO$_3$ as the prototype), and find 
that in oxygen-deficient
LiNbO$_{3-\delta}$, hydrostatic pressure induces an unexpected
metal-insulator transition between 8 and 9 GPa. 
Our calculations also find that
strong polar displacements persist in both metallic and insulating
oxygen-deficient LiNbO$_{3-\delta}$ and the size of polar
displacements is comparable to pristine LiNbO$_3$ under the same
pressure.  These properties are distinct from widely used perovskite
ferroelectric oxide BaTiO$_3$, whose polarization is quickly
suppressed by hydrostatic pressure and/or oxygen vacancies.  The
anomalous pressure-driven metal-insulator transition in
oxygen-deficient LiNbO$_{3-\delta}$ arises from the change of an
oxygen vacancy defect state.  Hydrostatic pressure increases the polar
displacements of oxygen-deficient LiNbO$_{3-\delta}$, which reduces
the band width of the defect state and eventually turns it into an
in-gap state.  In the insulating phase, the in-gap state is further
pushed away from the conduction band edge under hydrostatic pressure,
which increases the fundamental gap. Our work shows that for
LiNbO$_3$-type strong ferroelectrics, oxygen vacancies and hydrostatic
pressure combined can lead to new phenomena and potential functions,
in contrast to the harmful effects occurring to perovskite
ferroelectric oxides such as BaTiO$_3$.
\end{abstract}

\pacs{Valid PACS appear here}% PACS, the Physics and Astronomy
\maketitle

%\tableofcontents

\section{\label{sec:level1}Introduction}

Ferroelectricity is one of the most important functional properties of
transition metal
oxides~\cite{cohen1992origin,PhysRevLett.92.257201}. Two
classes of transition metal oxides: perovskite oxides (BaTiO$_3$ as
the prototype) and small $A$-site LiNbO$_3$-type oxides (LiNbO$_3$ as
the prototype) both exhibit robust ferroelectric polarization above
room
temperature~\cite{PhysRevB.53.1193,doi:10.1021/ja0758436,inaguma2008polar,
  shi2013ferroelectric}, promising potential applications in
ferroelectric-based electronic devices. 
The origin of ferroelectricity in BaTiO$_3$ arises from 
the hybridization between Ti-$d$ and 
O-$p$ states, which leads to a second-order Jahn-Teller distortion
(SOJT)~\cite{BERSUKER1966589,bersuker1978vibronic, doi:10.1021/ic50221a003}.
SOJT weakens the short-range repulsive forces that favor the 
non-polar structure and allows the long-range Coulombic forces to  
dominate, so that a polar structure is stabilized in BaTiO$_3$~\cite{cohen1992origin, Bersuker_2013}. 
On the other hand, 
ferroelectricity in LiNbO$_3$ originates from a geometric 
mechanism rather than charge transfer or
hybridization~\cite{PhysRevB.74.024102,doi:10.1021/jp402046t}. 
The very small Li ion size (thus a tolerance 
factor $t < 1$) leads to a combined structural distortions 
of NbO$_6$ octahedral rotation and a Li polar displacement, which 
altogether optimize the Li-O bond length~\cite{doi:10.1021/cm9801901,C5TC03856A}.
For BaTiO$_3$, it is known that either 
hydrostatic (positive) pressure~\cite{PhysRevB.74.180101,
doi:10.1063/1.3504194} or (charge neutral) oxygen 
vacancies~\cite{PhysRevB.84.064125,choi2011electronic,
PhysRevB.78.045107,PhysRevB.82.214109,PhysRevLett.104.147602} 
can reduce its polarization, thereby limiting 
its applications under those 
unfavorable conditions~\cite{PhysRevLett.109.247601,
PhysRevB.102.014108, D1TC01868J,ma2021large}. 
Therefore it is
interesting to explore how LiNbO$_3$ responds to oxygen vacancies
and/or hydrostatic pressure. In our previous study, we carefully
compare oxygen-deficient BaTiO$_{3-\delta}$ and LiNbO$_{3-\delta}$
\textit{under ambient pressure}~\cite{PhysRevMaterials.3.054405}. We
find that in BaTiO$_{3-\delta}$, the itinerant electrons doped by
oxygen vacancies are uniformly distributed, but in LiNbO$_{3-\delta}$,
the distribution of itinerant electrons is highly inhomogeneous.

In this work, we extend our study to hydrostatic pressure effects on
oxygen-deficient BaTiO$_{3-\delta}$ and LiNbO$_{3-\delta}$. We find a
more striking distinction between these two representative
ferroelectric materials. For BaTiO$_3$, since either charge neutral
oxygen vacancies
or hydrostatic pressure tends to suppress its ferroelectric property,
when these two factors are combined, expectedly it yields a more
unfavorable condition for sustaining its polarization. Our
first-principles calculations find that in oxygen-deficient
BaTiO$_{3-\delta}$, hydrostatic pressure can completely suppress its
polarization and drive it into a paraelectric state. However, for
LiNbO$_3$, in the presence of \textit{both} charge neutral oxygen 
vacancies and
external hydrostatic pressure, our calculations find an unexpected
pressure-driven metal-insulator transition 
between 8 and 9 GPa, in a wide
range of oxygen vacancy concentrations. The transition arises
from the change of an oxygen vacancy defect state in oxygen-deficient
LiNbO$_{3-\delta}$, which is mainly composed of Nb-$d$
orbitals. Applying a hydrostatic pressure increases 
the Nb-O polar displacements
in LiNbO$_{3-\delta}$, due to a cooperative
coupling between octahedral rotations and
polarity~\cite{PhysRevLett.120.197602}. This in turn 
reduces the band width of
the defect state. When the applied pressure makes the band width of
oxygen vacancy defect state sufficiently narrow in oxygen-deficient
LiNbO$_{3-\delta}$, the defect state splits off the
conduction bands and becomes an in-gap state, and thus the itinerant
electrons get trapped around the oxygen vacancy and a metal-insulator
transition occurs. Our calculations also find that in
both metallic and insulating oxygen-deficient LiNbO$_{3-\delta}$,
their polar displacements are  
comparable in magnitude to pristine LiNbO$_3$ under the same pressure.
These robust polar properties imply that LiNbO$_3$-based
ferroelectric materials can have a wider range of applications,
especially when they are under various conditions that are unfavorable
to perovskite ferroelectrics such as BaTiO$_3$.

%Our work shows that for LiNbO$_3$-type ferroelectrics, the presence of
%oxygen vacancies and applying hydrostatic pressure combined can lead
%to new phenomena and potential functions, in contrast to the negative
%effects occurring to perovskite ferroelectric oxides such as
%BaTiO$_3$. This implies that LiNbO$_3$-based
%ferroelectric devices can have a wider range of applications,
%especially when they are under various conditions that are unfavorable
%to BaTiO$_3$.

\section{\label{sce.level1}Computational details}

We perform density functional theory (DFT) calculations
~\cite{PhysRev.136.B864,PhysRev.140.A1133}, as implemented in Vienna
Ab-initio Simulation Package
(VASP)~\cite{RevModPhys.64.1045,PhysRevB.54.11169}. 
We use an energy cutoff of
600 eV. Charge self-consistent calculations are converged to 10$^{-5}$
eV. Both cell and internal coordinates are fully relaxed until each
force component is smaller than 10 meV/\AA\ and stress tensor is
smaller than 1 kbar. For the exchange-correlation functional,
we use local spin density approximation (LSDA)~\cite{PhysRevLett.45.566}.
However, we do not find
magnetization in either pristine BaTiO$_3$/LiNbO$_3$ or
oxygen-deficient BaTiO$_{3-\delta}$/LiNbO$_{3-\delta}$ 
(see Appendix~\ref{sec:LSDA}).
Therefore we sum over both spins in the calculation of density of states.
For pristine bulk calculations, we use a
tetragonal cell (5-atom) with a Monkhorst-Pack \textbf{k}-point
sampling of $12\times12\times12$ to study BaTiO$_3$ and find that $a$
= 3.94 \AA~and $c/a$ = 1.01; we use a hexagonal cell (30-atom) with a
Monkhorst-Pack \textbf{k}-point sampling of $10\times10\times10$ to
study $R3c$ LiNbO$_3$ and find that $a$ = 5.09 \AA~and $c$ = 13.80
\AA. Both of them are in good agreement with the previous
studies~\cite{PhysRevB.96.035143}. Bulk polarization is calculated
using Berry phase method~\cite{RevModPhys.66.899,resta2007theory,
SPALDIN20122}.
To study charge neutral oxygen
vacancies, we use supercell calculations.  For
  oxygen-deficient BaTiO$_{3-\delta}$, we start from the $P4mm$
  structure of pristine BaTiO$_3$ and remove one charge neutral oxygen
  atom. There are two inequivalent Wyckoff positions for oxygen atoms
  in the $P4mm$ structure~\cite{https://doi.org/10.1002/jcc.22942}.  
  We find similar results when oxygen
  vacancy is induced in either position. For oxygen-deficient
  LiNbO$_{3-\delta}$, we start from the $R3c$ structure of pristine
  LiNbO$_3$ and remove one charge neutral oxygen atom.
  There is only one Wyckoff position for oxygen atoms in
  the $R3c$ LiNbO$_3$, i.e.  all the oxygen atom positions are
  equivalent~\cite{PhysRevB.82.014104}. To simulate different oxygen
vacancy concentrations, we use supercells of different sizes in which
we remove one charge neutral oxygen vacancy. 
In the supercell calculations, we turn off all symmetries and 
fully relax the structure (both lattice constants and internal coordinates) 
to accommodate possible oxygen octahedral distortions and obtain the
ground state property.
We use a Monkhorst-Pack \textbf{k}-point
sampling of $8\times8\times8$ in supercell calculations. For the main
results, we use a 59-atom supercell for BaTiO$_{3-\delta}$ and
LiNbO$_{3-\delta}$ to simulate an oxygen vacancy concentration of
$\delta = 8.3\%$. We also use a 79-atom, 119-atom and 179-atom supercell of
LiNbO$_{3-\delta}$ to simulate a wide range of oxygen vacancy
concentrations ($\delta = 6.3\%, 4.2\%$ and $2.8\%$) and
test the robustness of our key results. 
We also use a higher energy cutoff
(750 eV) and a denser \textbf{k}-point sampling to test the key
results and we do not find any qualitative difference.

\section{Results}

For completeness and benchmarking, we first study hydrostatic pressure
effects on pristine BaTiO$_3$ and LiNbO$_3$. Then we compare
oxygen-deficient BaTiO$_{3-\delta}$ and LiNbO$_{3-\delta}$ under
hydrostatic pressure, and carefully study the
pressure-driven metal-insulator transition in oxygen-deficient
LiNbO$_{3-\delta}$.

\subsection{Pressure effects on pristine BaTiO$_3$ and LiNbO$_3$}

\begin{figure}[t!]
  \includegraphics[width=0.75\textwidth]{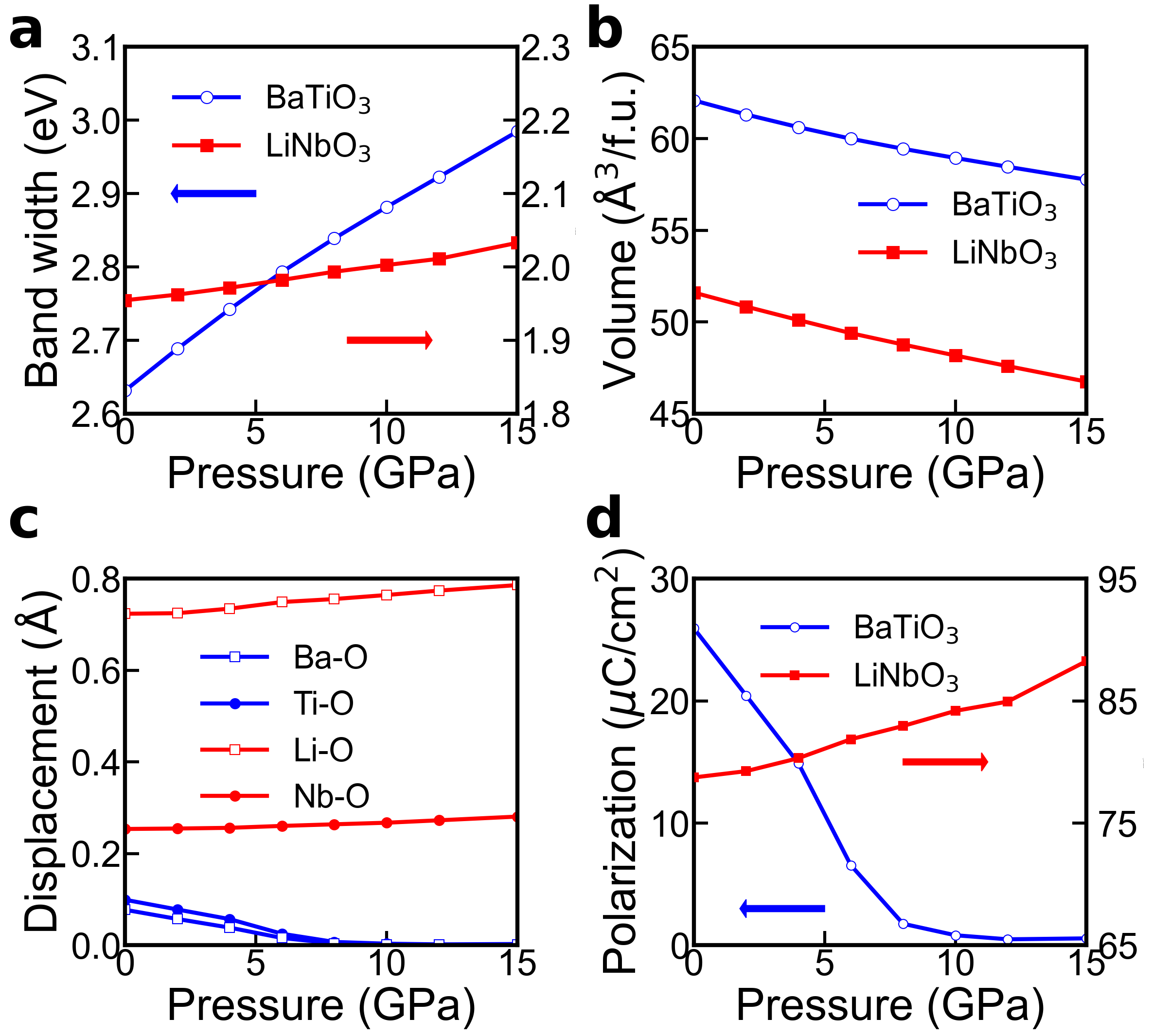}
  \caption{\label{fig:ferro-rigidband} Hydrostatic pressure effects on
    pristine BaTiO$_3$ and LiNbO$_3$. As a function of hydrostatic pressure,
    \textbf{a}) the band width of Ti-$t_{2g}$ states of BaTiO$_3$ and
    Nb-$t_{2g}$ states of LiNbO$_3$;
    \textbf{b}) the volume of BaTiO$_3$ and LiNbO$_3$ per formula;
    \textbf{c}) Ba-O and Ti-O polar displacements of BaTiO$_3$ and
                Li-O and Nb-O polar displacements of LiNbO$_3$;
    \textbf{d}) the polarization of BaTiO$_3$ and LiNbO$_3$.}
\end{figure}

We summarize the hydrostatic pressure effects on pristine BaTiO$_3$
and LiNbO$_3$ in Fig.~\ref{fig:ferro-rigidband}.  In both BaTiO$_3$
and LiNbO$_3$, the transition metal atoms (Ti and Nb) are in an oxygen
octahedron and thus the crystal field splitting removes the $d$
orbital degeneracy and separates the $d$ orbitals into two energy
groups~\cite{PhysRevB.71.245114}. The group with the lower (higher)
energy is called $t_{2g}$ ($e_g$) states. We focus on $t_{2g}$ states
for conciseness (we find similar results for $e_g$ states). Panel
\textbf{a} shows the band widths of Ti-$t_{2g}$ and Nb-$t_{2g}$ states
as a function of applied hydrostatic pressure (the electronic band
structures of pristine BaTiO$_3$ and LiNbO$_3$ under a representative
hydrostatic pressure are shown in Appendix~\ref{sec:elec}). 
Pressure naturally
decreases the volumes of BaTiO$_3$ and LiNbO$_3$ (see panel
\textbf{b}), which increases the hopping between Ti-$d$/Nb-$d$ orbitals
and O-$p$ orbitals.  As expected, the band width of both Ti-$t_{2g}$
and Nb-$t_{2g}$ states increases under hydrostatic pressure. However,
as panel \textbf{a} shows, the band width of Nb-$t_{2g}$ states
increases much more slowly than that of Ti-$t_{2g}$ states. To
understand that, we compare the polar displacements (the definition of
polar displacements in BaTiO$_3$ and LiNbO$_3$ can be found in
Appendix~\ref{sec:polar}) and the polarization of BaTiO$_3$ 
and LiNbO$_3$ under
hydrostatic pressure (panels \textbf{c} and \textbf{d}). Our
calculations find that hydrostatic pressure reduces the polar
displacements and polarization of BaTiO$_3$ and completely suppresses
them above a critical value, due to the destabilization of short-range
interaction at high pressure~\cite{PhysRevB.74.180101}. By contrast,
our calculations find that hydrostatic pressure increases the polar
displacements and polarization of LiNbO$_3$, due to a cooperative
coupling between octahedral rotations and
polarity~\cite{PhysRevLett.120.197602}. Similar
results have been obtained in ferroelectric 
ZnSnO$_3$~\cite{PhysRevLett.120.197602} and polar metal
LiOsO$_3$~\cite{doi:10.1063/1.5035133}. This
means that in LiNbO$_3$, there are two competing forces: the overall
volume reduction facilitates hopping, but the increased polar
displacements make hopping more difficult; while in BaTiO$_3$, both
volume reduction and weakened polar displacements help increase
hopping. Since band width is proportional to hopping, this explains
why the band width of Nb-$t_{2g}$ bands increases much more slowly than
that of Ti-$t_{2g}$ bands under hydrostatic pressure.

%\begin{figure}[t!]
%\includegraphics[width=0.9\textwidth]{primitive-bs-cbm}
%\caption{Band structures of pristine BaTiO$_3$ (\textbf{a1}) and
% LiNbO$_3$ (\textbf{a2}) near the conduction band minima
% (CBM). $t_{2g}$ states of Ti-$d$ and Nb-$d$ are highlighted by red
% color. }
%\label{fig:primitive-bs-cbm}
%\end{figure}
%\clearpage

\subsection{Pressure effects on oxygen-deficient BaTiO$_{3-\delta}$ 
and LiNbO$_{3-\delta}$}

\begin{figure}[t!]
\includegraphics[width=0.9\textwidth]{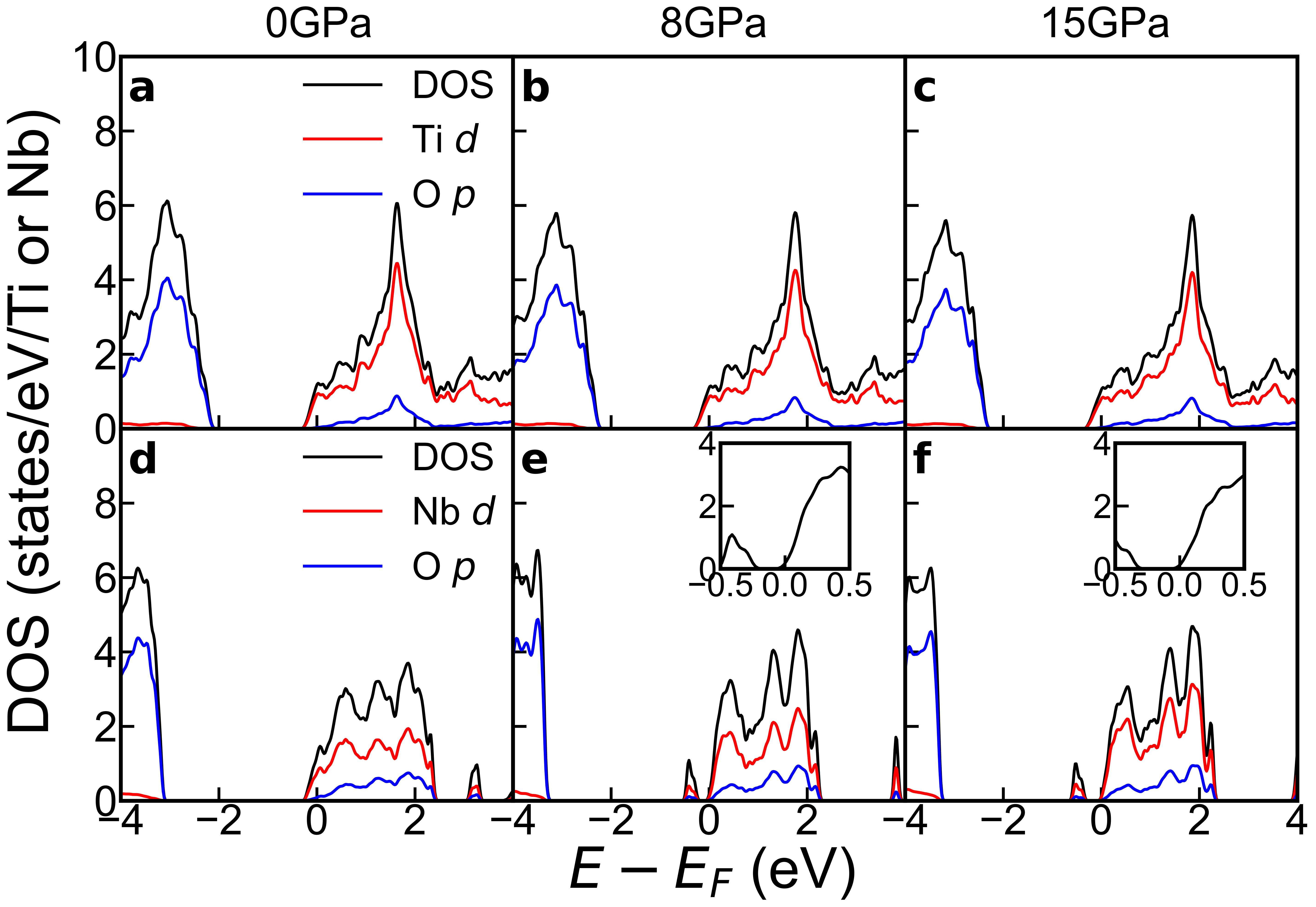}
\caption{ Densities of states of oxygen-deficient BaTiO$_{3-\delta}$
  (panels \textbf{a}, \textbf{b}, \textbf{c}) and oxygen-deficient 
  LiNbO$_{3-\delta}$ (panels
  \textbf{d}, \textbf{e}, \textbf{f}) with $\delta = 8.3\%$ 
  under a hydrostatic pressure of 0,
  8 and 15 GPa, respectively. The black, red and blue are total,
  Ti-$d$ or Nb-$d$ and O-$p$ projected densities of states,
  respectively. The insets in \textbf{e} and \textbf{f} show the
  densities of states near the Fermi level. }
\label{fig:59-PDOS}
\end{figure}

Next we study hydrostatic pressure effects on oxygen-deficient
BaTiO$_{3-\delta}$ and oxygen-deficient LiNbO$_{3-\delta}$ and show
that under pressure the electronic structure of these two materials
behaves much more differently than pristine ones. Charge neutral oxygen
vacancies are a common defect in 
complex oxides~\cite{PhysRevB.84.245206, PhysRevB.86.161102, 
PhysRevB.88.054111, zhou2021ab,
RevModPhys-Walle}, which are electron
donors and induce defect states that are close to conduction band
edge~\cite{PhysRevB.90.085202,C8NR09666J,doi:10.1063/1.5143309}. If
the band width of the defect states is large enough so that the Fermi
level cuts through both defect bands and conduction bands, then the
system is conducting and the doped electrons are itinerant.

Fig.~\ref{fig:59-PDOS} shows the density of states (DOS) of
oxygen-deficient BaTiO$_{3-\delta}$ and LiNbO$_{3-\delta}$ (with
$\delta = 8.3\%$) as a function of applied hydrostatic pressure.  We
find that under hydrostatic pressure, oxygen-deficient
BaTiO$_{3-\delta}$ is conducting and its DOS almost does not change
under increasing pressure (panels \textbf{a}, \textbf{b},
\textbf{c}).  By contrast, our calculations find that
LiNbO$_{3-\delta}$ undergoes a metal-insulator transition. When the
applied hydrostatic pressure is smaller than 8 GPa, LiNbO$_{3-\delta}$
is conducting; and above 8 GPa, a small gap is opened in the
electronic structure of LiNbO$_{3-\delta}$ and the gap is further
increased with pressure (see the insets of panels \textbf{e} and
\textbf{f}).  For pristine materials, this pressure-driven
metal-insulator transition is rare, because pressure generically
decreases volume and usually closes gap and makes materials more
conducting.

\begin{figure}[t!]
\includegraphics[width=0.85\textwidth]{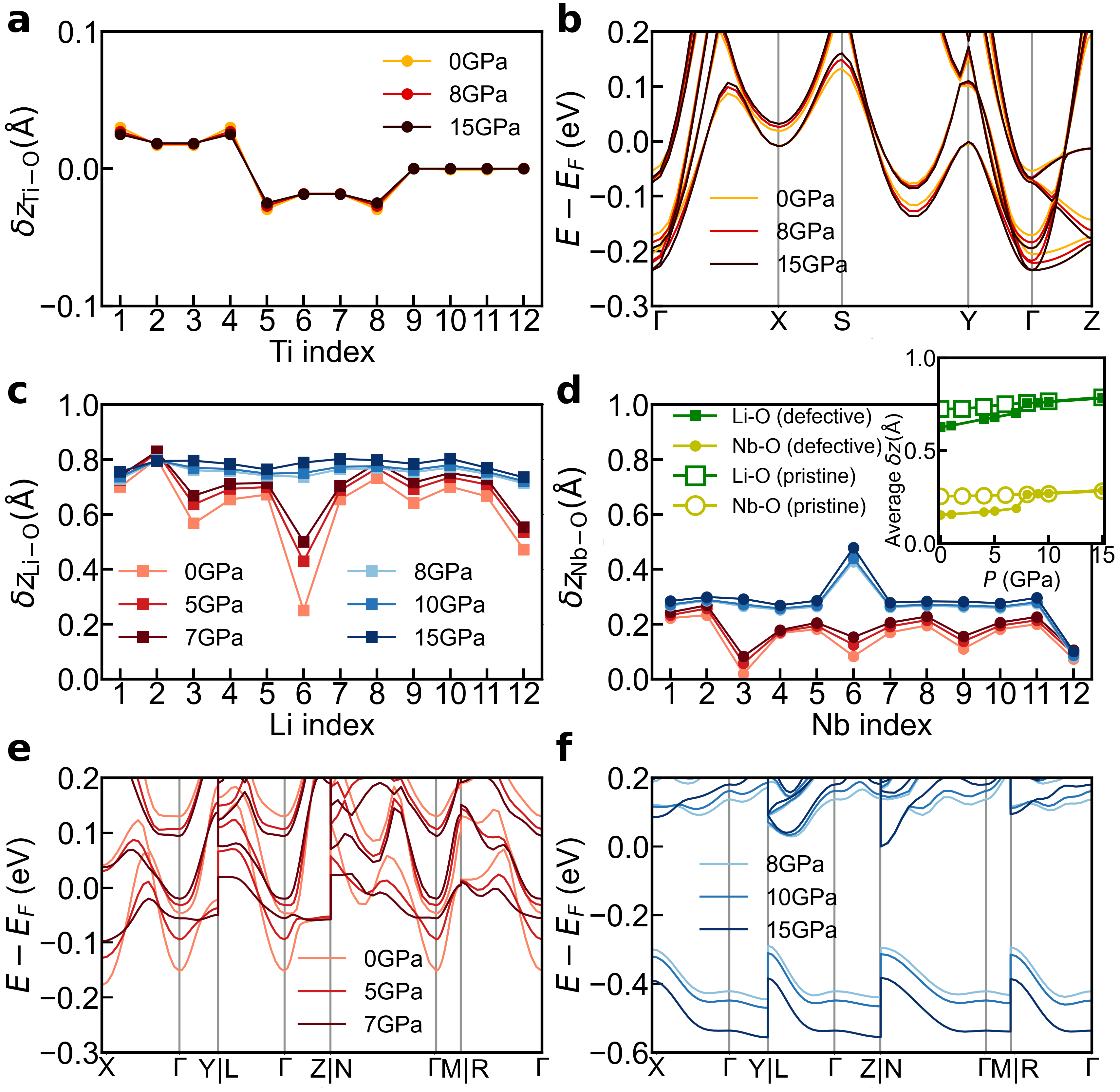}
\caption{ Panels \textbf{a}-\textbf{b}: oxygen-deficient
  BaTiO$_{3-\delta}$ (with $\delta = 8.3\%$) under a hydrostatic
  pressure of 0, 8 and 15 GPa.  \textbf{a}: polar displacements of
  each Ti atom. \textbf{b}: electronic band structure close to the
  Fermi level. The high-symmetry \textbf{k}-path is
  $\Gamma$(0,0,0)-X(0.5,0,0)-S(0.5,0.5,0)-Y(0,0.5,0)-$\Gamma$(0,0,0)-
  Z(0,0,0.5).  Panels \textbf{c}-\textbf{f}: oxygen-deficient
  LiNbO$_{3-\delta}$ (with $\delta = 8.3\%$) under several hydrostatic
  pressures in the range of 0-15 GPa. \textbf{c}: polar displacements of each Li
  atom. \textbf{d}: polar displacements of each Nb atom. Red and blue
  squares represent conducting and insulating phases of
  oxygen-deficient LiNbO$_{3-\delta}$, respectively. The
  inset in \textbf{d} compares the average polar displacements of Li and Nb
  in oxygen-deficient LiNbO$_{3-\delta}$ (with $\delta = 8.3\%$) and
  pristine LiNbO$_3$.
  \textbf{e}: electronic band structure close to the Fermi level under
  a hydrostatic pressure of 0, 5 and 7 GPa, where LiNbO$_{3-\delta}$
  is conducting. \textbf{f}: electronic band structure close to the
  Fermi level under a hydrostatic pressure of 8, 10 and 15 GPa, where
  LiNbO$_{3-\delta}$ is insulating.  The high-symmetry
  \textbf{k}-point path is
  X(0,-0.5,0)-$\Gamma$(0,0,0)-Y(0.5,0,0)$
  \mid$L(0.5,-0.5,0)-$\Gamma$(0,0,0)-Z(-0.5,0,0.5)$
  \mid$N(-0.5,-0.5,0.5)-$\Gamma$(0,0,0)-M(0,0,0.5)$\mid$R(0,-0.5,0.5)-
  $\Gamma$(0,0,0).}
\label{fig:59-dis}
\end{figure}

To understand the origin of this pressure-driven metal-insulator
transition, we study polar displacements (the definition of
polar displacements in BaTiO$_{3-\delta}$ and LiNbO$_{3-\delta}$
is also found in Appendix~\ref{sec:polar}) and band structure of
oxygen-deficient LiNbO$_{3-\delta}$ and compare them to those of
oxygen-deficient BaTiO$_{3-\delta}$ (with $\delta=8.3\%$).
Fig.~\ref{fig:59-dis}\textbf{a} shows the site-resolved Ti-O polar
displacements in BaTiO$_{3-\delta}$ under hydrostatic pressure. 
Different from
pristine BaTiO$_3$, each oxygen vacancy donates two mobile electrons,
which already suppress the average Ti-O polar displacements under
ambient pressure. Applying a hydrostatic pressure does not affect the Ti-O
polar displacements. Fig.~\ref{fig:59-dis}\textbf{b} shows the electronic
band structure of oxygen-deficient BaTiO$_{3-\delta}$. The lowest band that
crosses the Fermi level is an oxygen vacancy defect state.
Applying a hydrostatic pressure only slightly increases the band width 
of the
defect state and has little effect on the overall electronic structure.
Fig.~\ref{fig:59-dis}\textbf{c} and \textbf{d} show the
site-resolved Li-O and Nb-O polar displacements in oxygen-deficient
LiNbO$_{3-\delta}$ under hydrostatic pressure. The red (blue) squares in
Fig.~\ref{fig:59-dis}\textbf{c} and \textbf{d} refer to the polar
displacements in the conducting (insulating) LiNbO$_{3-\delta}$. In the
inset of panel \textbf{d}, we compare the average Li-O and
Nb-O polar displacements between oxygen-deficient LiNbO$_{3-\delta}$ and
pristine LiNbO$_3$ under hydrostatic pressure.
We find that similar to pristine LiNbO$_3$ 
(see Fig.~\ref{fig:ferro-rigidband}),
hydrostatic pressure increases the polar displacements of oxygen-deficient
LiNbO$_{3-\delta}$. In particular, 
the Nb atom and the Li atom that are closest to the oxygen vacancy 
have the most 
substantial increase in the polar displacement.
However, different from pristine LiNbO$_3$, there
is a sudden ``jump'' in the polar displacements between 7 and 8 GPa,
which is evident in the inset of panel \textbf{d}.
For a given hydrostatic pressure, the average polar 
displacements of conducting
oxygen-deficient LiNbO$_{3-\delta}$ are slightly smaller 
than those of pristine
LiNbO$_{3}$. This is because mobile electrons in conducting
LiNbO$_{3-\delta}$ can screen internal
electric fields and suppress the polar 
displacements~\cite{PhysRevMaterials.3.054405}. However, in 
insulating LiNbO$_{3-\delta}$, the average polar displacements
are almost the same as those of pristine LiNbO$_3$
because screening from mobile electrons is absent.

Accompanying this sudden ``jump'' in the polar displacements is a
metal-insulator transition. Fig.~\ref{fig:59-dis}\textbf{e} and
\textbf{f} show the electronic band structures of conducting and insulating
oxygen-deficient LiNbO$_{3-\delta}$, respectively.  In conducting
LiNbO$_{3-\delta}$, the band width of the defect state is reduced by
hydrostatic pressure, which is in
contrast to pristine LiNbO$_3$ and BaTiO$_3$ in which hydrostatic pressure
always increases their band width. This anomalous pressure effect is
because the defect state is
closely related to Nb \#6, which is the nearest neighbor of the oxygen
vacancy. The substantially increased polar displacement of Nb \#6 makes
the electron hopping much more difficult, which overweighs the volume
reduction from hydrostatic pressure. In insulating LiNbO$_{3-\delta}$, 
the defect
state becomes an in-gap state.  Similar to defects in semiconductors,
hydrostatic pressure increases the repulsion between conduction band 
edge and
in-gap states, which further increases the fundamental band gap of
oxygen-deficient LiNbO$_{3-\delta}$ (i.e. the gap between the 
conduction band
edge and the defect state)~\cite{PhysRevB.25.7661,
doi:10.1063/1.1525045,PhysRevB.78.235104,PhysRevB.74.165204,
PhysRevLett.115.126806}.

\begin{figure}[t]
\includegraphics[width=0.95\textwidth]{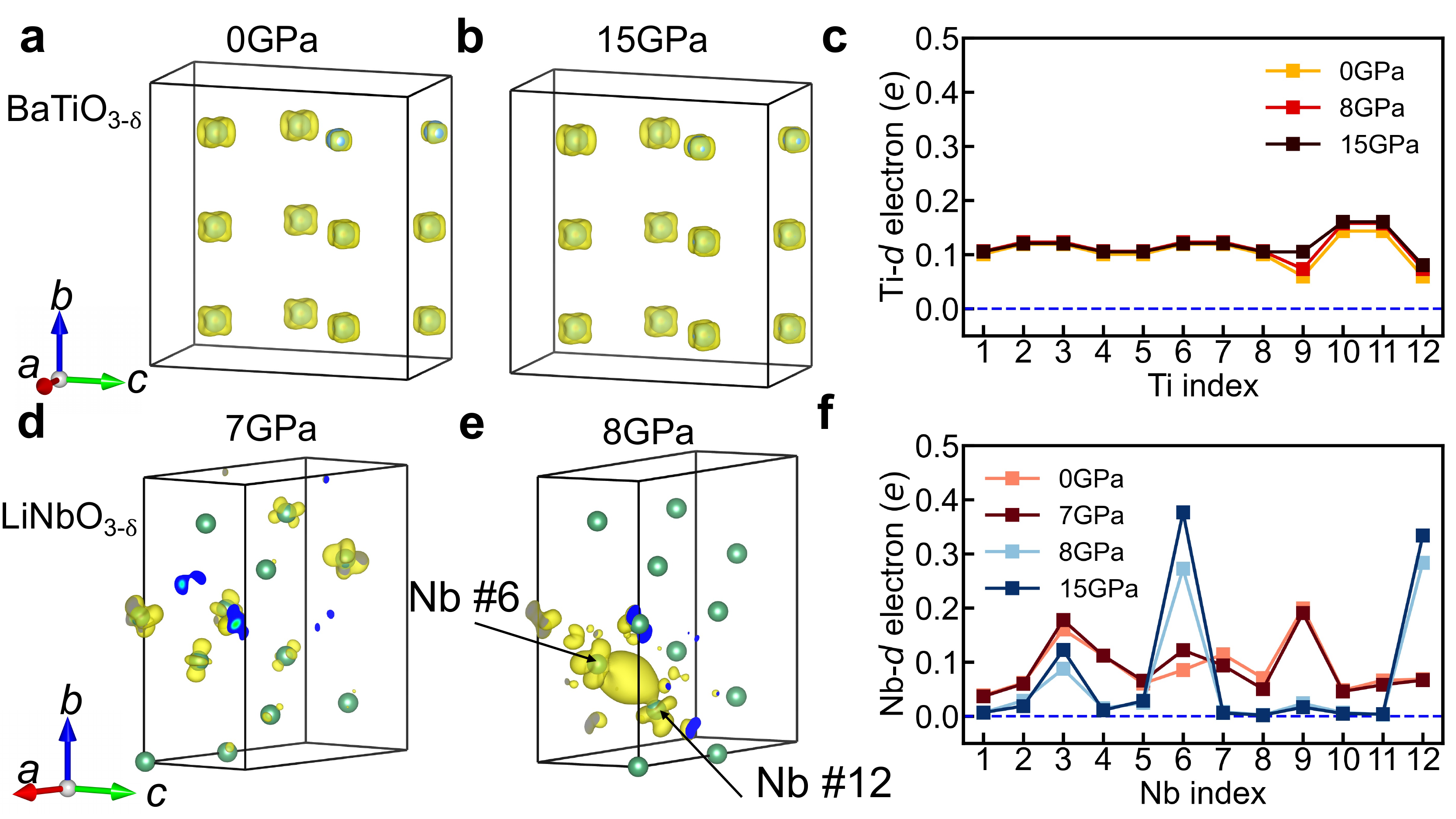}
\caption{Panels \textbf{a} and \textbf{b} show the iso-value surfaces
  of doped electron distribution in oxygen-deficient
  BaTiO$_{3-\delta}$ with $\delta = 8.3\%$ under 0 and 15GPa,
  respectively. Panel \textbf{c} shows the doped electrons on each Ti
  atom in oxygen-deficient BaTiO$_{3-\delta}$ ($\delta = 8.3\%$) under
  a hydrostatic pressure of 0, 8 and 15 GPa. Panels \textbf{d} and
  \textbf{e} are the iso-value surfaces of doped electron distribution
  in oxygen-deficient LiNbO$_{3-\delta}$ with $\delta = 8.3\%$ under 7
  and 8 GPa, respectively. Panel \textbf{f} shows the doped electrons
  on each Nb atom in oxygen-deficient LiNbO$_{3-\delta}$ ($\delta =
  8.3\%$) under a hydrostatic pressure of 0, 7, 8 and 15 GPa.  For
  both materials, the iso-surface corresponds to a charge density of
  0.027$e$/\AA$^{3}$.  The number of doped electrons on each Ti or Nb
  atom is obtained by integrating their $d$ states from the valence
  band edge to the Fermi level.}
\label{fig:FIG4}
\end{figure}

Next we study how hydrostatic pressure changes the spatial
distribution of doped electrons in 
oxygen-deficient BaTiO$_{3-\delta}$ and oxygen-deficient
LiNbO$_{3-\delta}$. Each oxygen vacancy donates two electrons and they
occupy the defect state and conduction bands, which are mainly
composed of Ti-$d$ or Nb-$d$ states.  Fig.~\ref{fig:FIG4} shows the
spatial distribution of doped electrons and the number of doped
electrons on each Ti and Nb atoms in oxygen-deficient
BaTiO$_{3-\delta}$ and LiNbO$_{3-\delta}$ (with $\delta =
8.3\%$). Oxygen-deficient BaTiO$_{3-\delta}$ is always conducting and
the doped electrons are almost homogeneously distributed, as is shown
explicitly in Fig.~\ref{fig:FIG4}\textbf{a} and
\textbf{b}. Hydrostatic pressure makes the electron distribution even
more homogeneous in BaTiO$_{3-\delta}$ (see
Fig.~\ref{fig:FIG4}\textbf{c}). However, in oxygen-deficient
LiNbO$_{3-\delta}$, the distribution of doped electrons is highly
inhomogeneous~\cite{PhysRevMaterials.3.054405}. As we showed
previously in Fig.~\ref{fig:59-dis}, hydrostatic pressure increases
the polar displacements of LiNbO$_{3-\delta}$ and narrows the band
width of the defect state, which means the reduction of electron
hopping. Correspondingly, the doped electrons tend to be localized in
real space. From ambient pressure to 7 GPa, the doped electrons are
distributed on all Nb sites (see Fig.~\ref{fig:FIG4}\textbf{d}).
However, under a hydrostatic pressure of 8 GPa that is just above the
critical pressure, most doped electrons are concentrated in the void
space between Nb \#6 and Nb \#12, which are the two nearest neighbors
of the oxygen vacancy (see Fig.~\ref{fig:FIG4}\textbf{e}). Such a
localization effect can be more clearly seen from the two peaks in the
Nb site-resolved electron distribution, shown in
Fig.~\ref{fig:FIG4}\textbf{f}. In oxygen-deficient LiNbO$_{3-\delta}$,
applying hydrostatic pressure localizes itinerant electrons and leads
to a metal-insulator transition, in stark contrast to oxygen-deficient
BaTiO$_{3-\delta}$. We note that 
the localized carriers in the insulating LiNbO$_{3-\delta}$
(Fig.~\ref{fig:FIG4}\textbf{e}) 
correspond to \textit{two} electrons that are donated by an oxygen 
vacancy. These two electrons completely fill the defect 
band (see Fig.~\ref{fig:59-dis}\textbf{f}) and thus 
spin up and spin down channels
have equal occupancy. Therefore,  
no local magnetic moment emerges from this charge localization. 
Our case is different from a recent work on doped BaTiO$_3$ 
under epitaxial strain~\cite{xu2019electron} in which \textit{one} electron
half-fills a band and occupies a single spin channel.  
Correlation effects localize the lone electron and 
a magnetic polaron state is formed~\cite{Mott_1949}.

\begin{figure}[t!]
\includegraphics[width=0.85\textwidth]{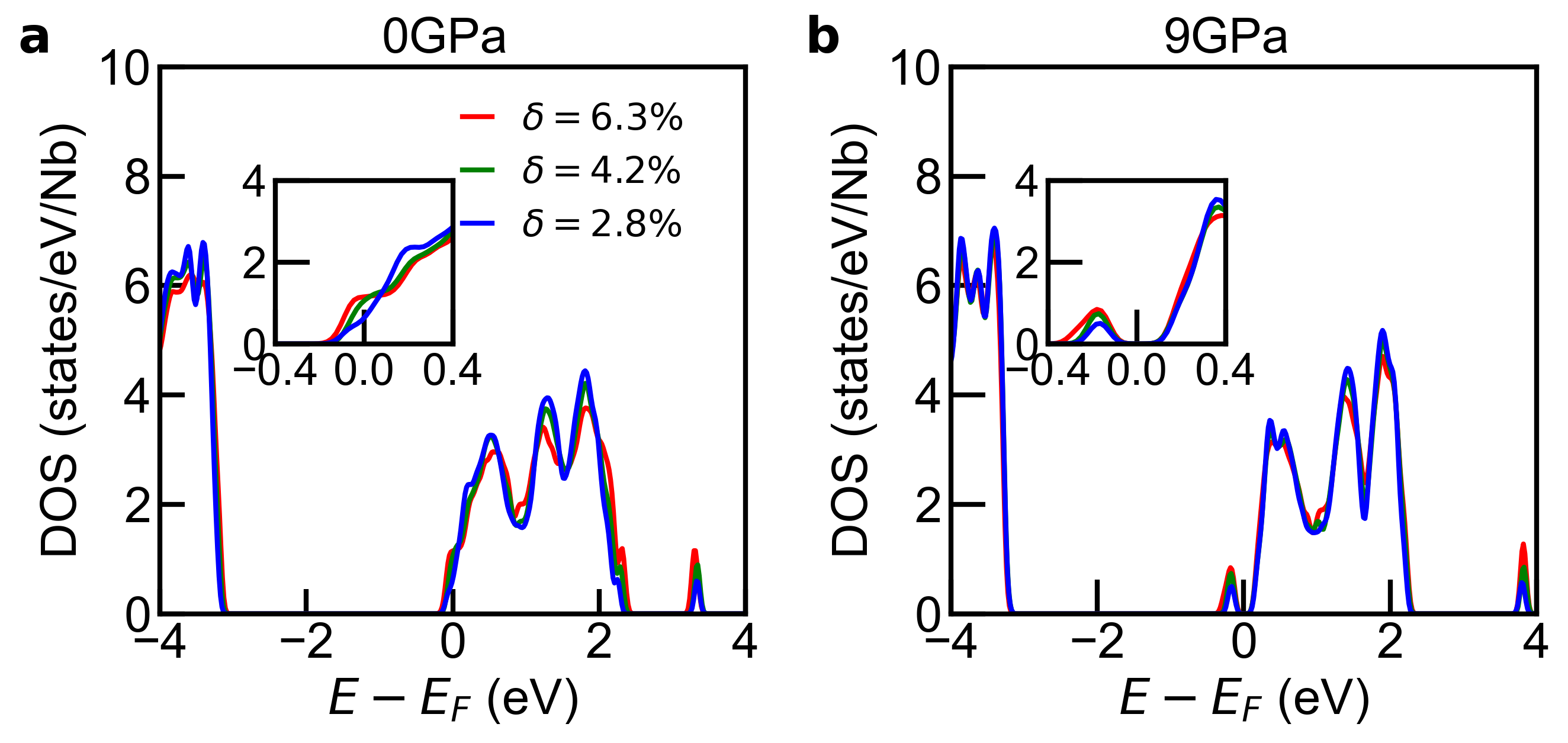}
\caption{Densities of states of oxygen-deficient LiNbO$_{3-\delta}$
  with different vacancy concentrations
  under 0 (panel \textbf{a}) and 9 GPa (panel \textbf{b}). The red,
  green and blue curves are for LiNbO$_{3-\delta}$ with $\delta =
  6.3\%$, $\delta = 4.2\%$ and $\delta = 2.8\%$, respectively. The
  insets show the densities of states near the Fermi level.}
\label{fig:FIG5}
\end{figure}

So far, we have only studied one concentration of oxygen vacancy
$\delta = 8.3\%$.  In Fig.~\ref{fig:FIG5} we study different oxygen
vacancy concentrations in oxygen-deficient LiNbO$_{3-\delta}$. 
We find that the
pressure-driven metal-insulator transition occurs in a wide range of
vacancy concentrations. In addition to $\delta = 8.3\%$ (59-atom
cell), we also calculate $\delta = 6.3\%$ (79-atom cell), $\delta =
4.2\%$ (119-atom) and $2.8\%$ (179-atom cell). 
In all three cases,
under ambient pressure, LiNbO$_{3-\delta}$ is conducting
(Fig.~\ref{fig:FIG5}\textbf{a}) while under a hydrostatic pressure of 
9 GPa, LiNbO$_{3-\delta}$ becomes
insulating with the defect state turning into an in-gap state and
lying below the conduction bands (Fig.~\ref{fig:FIG5}\textbf{b}).  The
critical pressure weakly depends on vacancy concentration, which ranges
between 8 and 9 GPa for oxygen-deficient LiNbO$_{3-\delta}$.

Finally, we make two comments. 
One is that the pressure-driven 
metal-insulator transition in our study is fundamentally different from 
the pressure-driven insulator-to-metal transition in 
correlated materials~\cite{kunevs2008collapse,PhysRevLett.102.146402,PhysRevB.92.035125,Huang_2017}.
In the former, high pressure stabilizes 
an \textit{insulating} state in oxygen-deficient LiNbO$_{3-\delta}$; 
while in the latter, increasing pressure closes the Mott gap and 
leads to a \textit{metallic} state. The pressure-driven insulator-to-metal 
transition in correlated materials is usually strongly first-order, 
accompanied by abrupt reduction in volume and/or collapse of local 
magnetic moment~\cite{PhysRevLett.94.115502,PhysRevX.8.031059, PhysRevB.101.245144}. By contrast, the pressure-driven 
metal-insulator transition in our study is associated with 
a defect band, which is fully occupied when the fundamental gap 
is opened. As we mentioned above, since the defect band is 
fully occupied (in both spin up and 
spin down channels), no local magnetic moment is found in 
our calculations and correlation effects play a minor role.
The second comment is the connection between our results and
available experimental data.Oxygen vacancies have been 
widely found in important ferroelectric oxides, such as BaTiO$_3$~\cite{doi:10.1063/1.4936159}, 
PbTiO$_3$~\cite{nishida2013oxygen}, (Pb$_{1-x}$Ba$_x$)(Zr$_{0.95}$Ti$_{0.05}$)O$_3$~\cite{doi:10.1063/1.4900610,PhysRevB.98.094102}, 
KNbO$_3$~\cite{doi:10.1063/1.3309745}, LiNbO$_3$~\cite{doi:10.1063/1.3693041}, HfO$_2$~\cite{doi:10.1063/1.4940370,Strand_2018}, and amorphous Al$_2$O$_3$~\cite{peng2020ferroelectric}. 
The presence of oxygen vacancies can substantially change the 
structural, dielectric, and transport properties of 
ferroelectrics~\cite{PhysRevLett.109.247601, ma2021large}, such as the reduction of polarization~\cite{doi:10.1063/5.0023554,peng2021oxygen} and the 
emergence of conduction~\cite{doi:10.1063/1.3309745,doi:10.1063/1.4900610}. Sometimes combining oxygen vacancies and 
polar structure in ferroelectrics can also lead to new functions~\cite{noguchi2019ferroelectrics,peng2020ferroelectric,yu2021synergy}.
Most pertinent to the current study are oxygen-deficient 
BaTiO$_{3-\delta}$ and LiNbO$_{3-\delta}$.
Oxygen vacancies in BaTiO$_3$ have been observed and 
studied in a few
experiments~\cite{PhysRevB.78.045107,
PhysRevB.82.214109,PhysRevLett.104.147602}. The
theoretical results in the current study are
qualitatively consistent with the experiments in that 1) conduction
appears in BaTiO$_{3-\delta}$ due to oxygen vacancies~\cite{PhysRevB.78.045107,PhysRevB.82.214109}; 2) Ti-O polar
displacements are reduced by itinerant electrons
and are completely suppressed above a
critical electron concentration of $1.9 \times 10^{21}$ cm$^{-3}$~\cite{
PhysRevLett.104.147602}.
For LiNbO$_3$, various types of defects such as Li-vacancies,
O-vacancies, Li-Nb antisite defects, etc. are found in
experiments~\cite{bredikhin2000nonstoichiometry,
doi:10.1063/1.1536958,karapetyan2006thermal,
lengyel2015growth}. In
particular, oxygen-deficient LiNbO$_{3-\delta}$ in nanocrystallites
and single crystals have been reported in 
several experiments~\cite{DIAZMORENO201482,MANIKANDAN2015156,
doi:10.1063/1.3673434,doi:10.1111/jace.16522},
but high-pressure study has not been performed. We hope that our
theoretical study may stimulate further experiments on
oxygen-deficient LiNbO$_{3-\delta}$ under hydrostatic pressure.

\clearpage
\section{Conclusion}

In summary, we compare the pressure effects on two important and
representative ferroelectric oxides BaTiO$_3$ and LiNbO$_3$ in the
pristine form and in a defective form with oxygen vacancies. In
pristine BaTiO$_3$, hydrostatic pressure reduces the polar displacements and
increases the band width of Ti-$d$ states. In oxygen-deficient
BaTiO$_{3-\delta}$, pressure makes the doped electrons more
homogeneously distributed. Both phenomena are within expectation. By
contrast, in both pristineLiNbO$_{3}$ and oxygen-deficient
LiNbO$_{3-\delta}$, pressure increases their
polar displacements. More strikingly, 
in oxygen-deficient LiNbO$_{3-\delta}$,
we find an unexpected pressure-driven metal-insulator transition.
The anomalous transition arises from the fact that 
the substantially increased polar displacements in oxygen-deficient
LiNbO$_{3-\delta}$ under hydrostatic pressure
reduce the overlap between Nb-$d$ and O-$p$
orbitals and thus decrease the band width of the defect state. When
the band width of the defect state is sufficiently narrow, it turns
into an in-gap state and the system becomes insulating with itinerant
electrons trapped around the oxygen vacancy.
This pressure-driven metal-insulator
transition occurs to oxygen-deficient LiNbO$_{3-\delta}$ in a wide range of
oxygen vacancy concentrations.

Our work shows that LiNbO$_3$-type ferroelectric materials have more
robust polar properties against oxygen vacancies and hydrostatic pressure
than BaTiO$_3$. Furthermore, the intriguing 
pressure-driven metal-insulator
transition in oxygen-deficient LiNbO$_{3-\delta}$ is not found
in widely used ferroelectric materials BaTiO$_3$
or BaTiO$_{3-\delta}$.
This implies that LiNbO$_3$-based
ferroelectric devices may have a wider range of applications, in
particular when they are under conditions that are unfavorable to
perovskite ferroelectrics such as BaTiO$_3$.

\begin{acknowledgments}
  Hanghui Chen is supported by the National Natural Science Foundation
  of China under project number 11774236 and NYU University Research
  Challenge Fund. Yue Chen and Chengliang Xia are supported by the
  Research Grants Council of Hong Kong under Project Number
  17201019. The authors are grateful for the research computing
  facilities offered by ITS, HKU and high-performance computing
  resources from NYU Shanghai.
\end{acknowledgments}

\appendix

\newpage
\clearpage

\section{\label{sec:LSDA}LSDA calculations for possible magnetization}

In this section, we show the LSDA calculations of pristine 
BaTiO$_3$ and LiNbO$_3$, as well as oxygen-deficient 
BaTiO$_{3-\delta}$
and LiNbO$_{3-\delta}$. We find that while we intentionally 
break the spin symmetry, the resulting 
density of states do not exhibit any magnetization in our 
calculations.

\begin{figure}[b]
\includegraphics[width=0.8\textwidth]{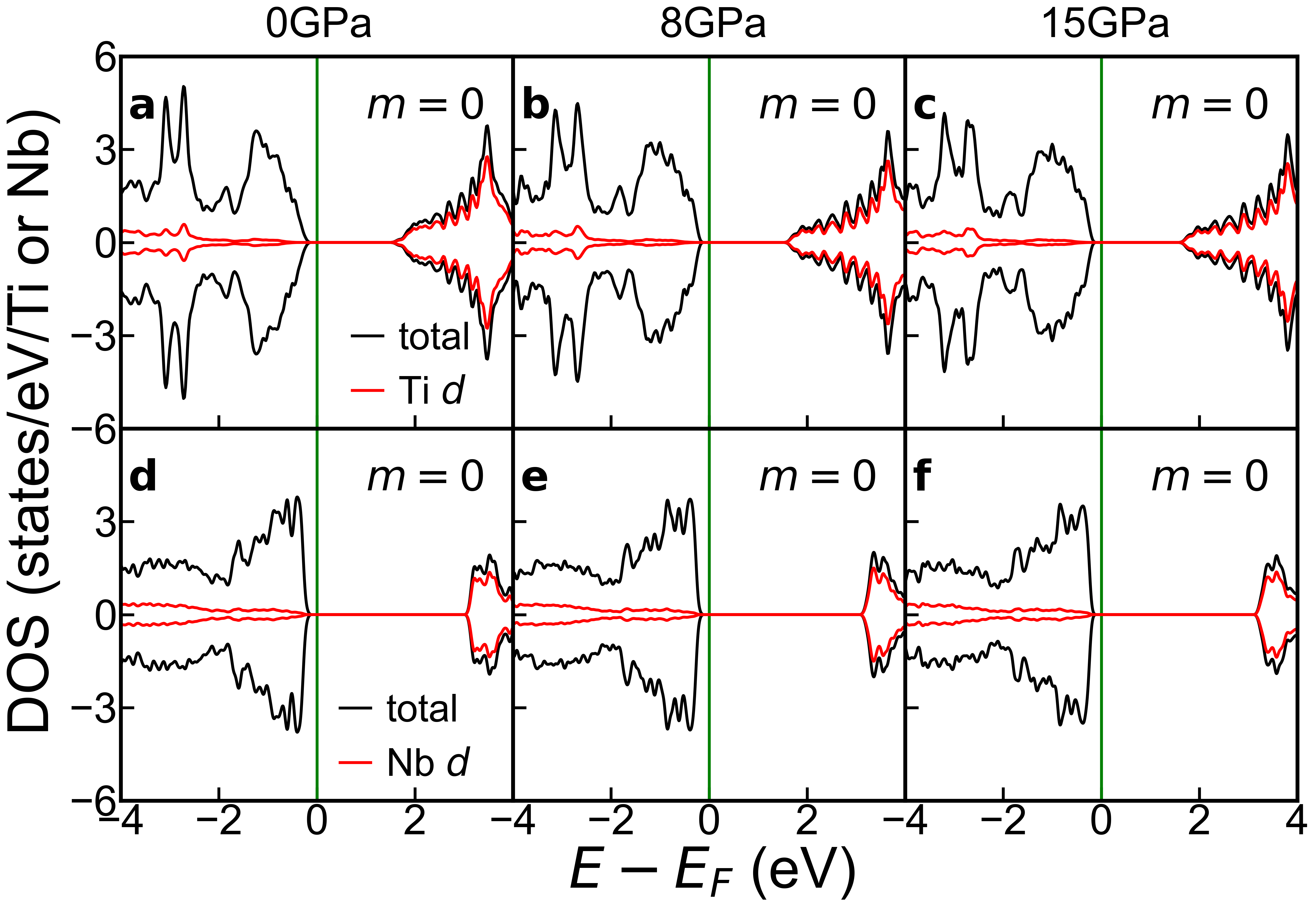}
\caption{Densities of states of pristine 
BaTiO$_{3}$ (panels \textbf{a}, \textbf{b}, \textbf{c}) 
and pristine LiNbO$_{3}$ 
(panels \textbf{d}, \textbf{e}, \textbf{f}) 
 under a hydrostatic pressure of 
0, 8 and 15 GPa. The black and red lines 
are total and Ti-$d$/Nb-$d$ projected densities of 
states, respectively. 
$m$ shown in each sub-plot is the average magnetization per 
Ti or per Nb. The green solid line is the Fermi level.}
\label{fig:pristine-DOS-spin}
\end{figure}

\begin{figure}[ht!]
\includegraphics[width=0.8\textwidth]{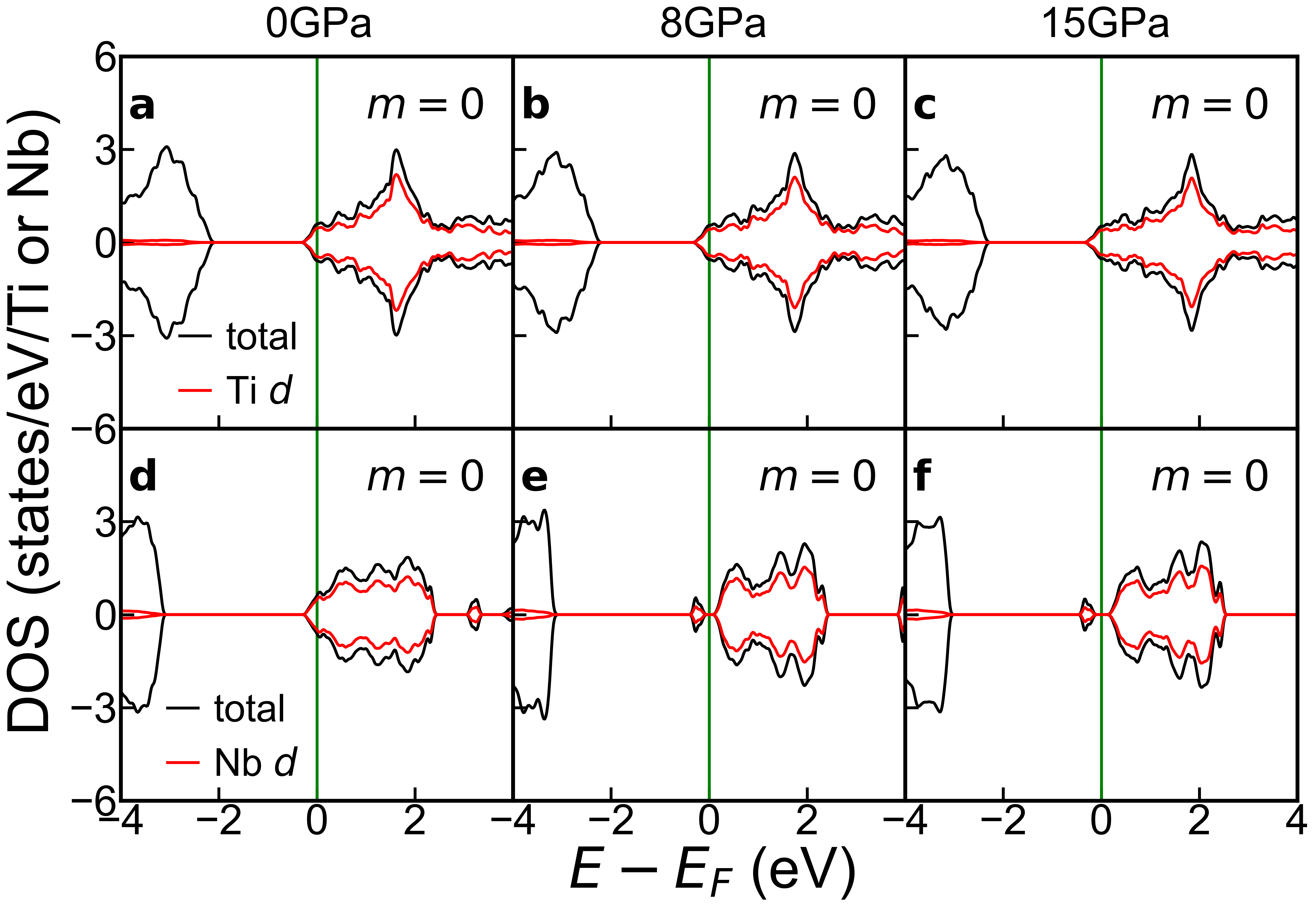}
\caption{Densities of states of oxygen-deficient 
BaTiO$_{3-\delta}$ (panels \textbf{a}, \textbf{b}, \textbf{c}) 
and oxygen-deficient LiNbO$_{3-\delta}$ 
(panels \textbf{d}, \textbf{e}, \textbf{f}) 
with $\delta$ = 8.3 \% under a hydrostatic pressure of 
0, 8 and 15 GPa. The black and red lines 
are total and Ti-$d$/Nb-$d$ projected densities of 
states, respectively. 
$m$ shown in each sub-plot is the average magnetization per 
Ti or per Nb. The green solid line is the Fermi level.}
\label{fig:59-DOS-spin}
\end{figure}

Fig.~\ref{fig:pristine-DOS-spin} shows the density 
of states of pristine BaTiO$_3$ and 
LiNbO$_3$ under a hydrostatic pressure of 0, 8 and 15 GPa. Fig.~\ref{fig:59-DOS-spin} shows the density of states of 
oxygen-deficient BaTiO$_{3-\delta}$ and 
LiNbO$_{3-\delta}$ under a hydrostatic pressure of 0, 8 and 15 GPa.

\newpage
\clearpage

\section{\label{sec:elec}Electronic band structure of pristine 
BaTiO$_3$ and LiNbO$_3$ under a hydrostatic pressure}

\begin{figure}[t!]
\includegraphics[width=0.7\textwidth]{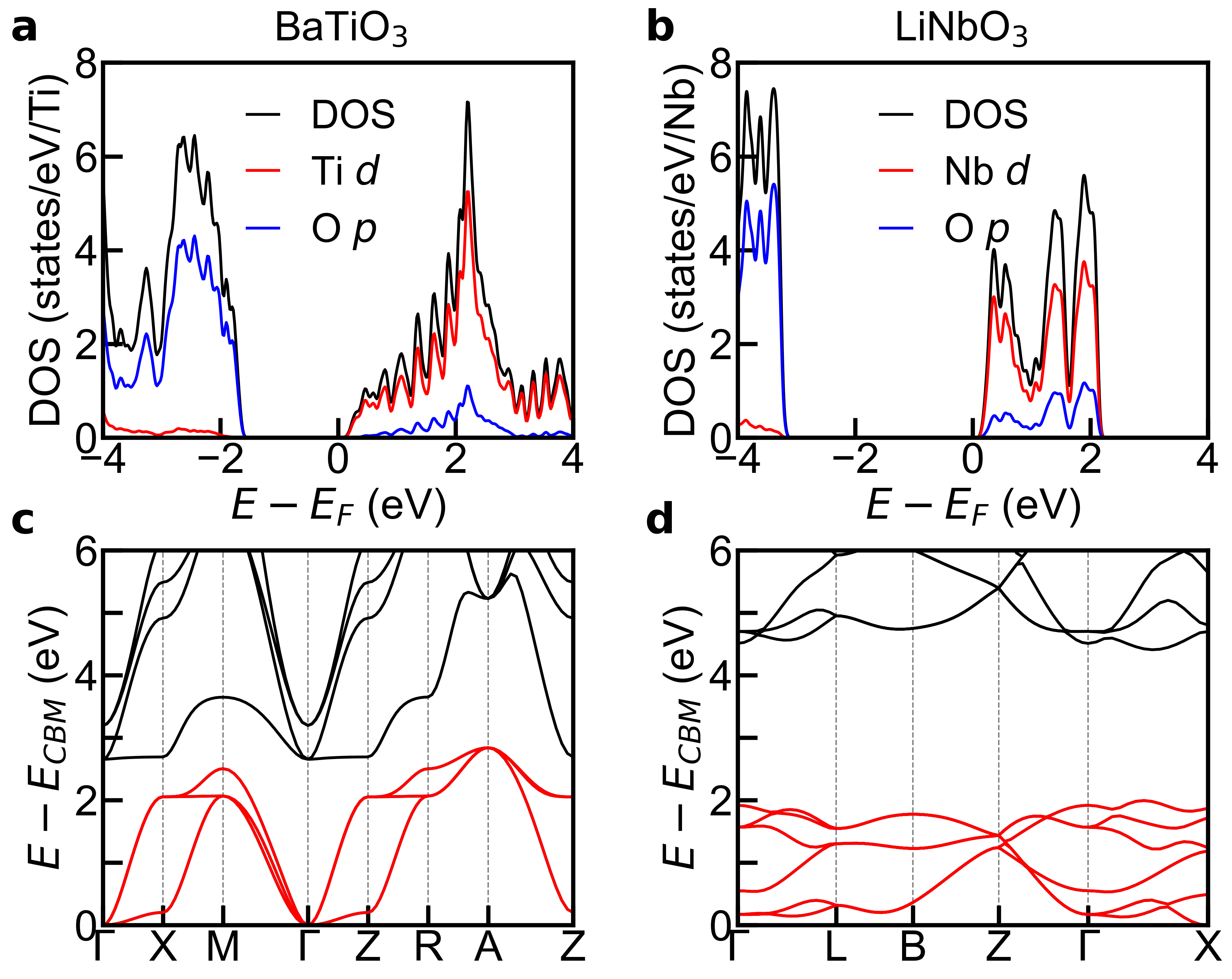}
\caption{Densities of states and electronic 
   band structure of pristine BaTiO$_3$ (panels \textbf{a} 
   and \textbf{c}) and pristine LiNbO$_3$ (panels \textbf{b} 
   and \textbf{d}) under a hydrostatic pressure of 8 GPa. 
   The black, red, and blue lines in \textbf{a} and \textbf{b} 
   are total, Ti-$d$ or Nb-$d$, and O-$p$ projected densities 
   of states, respectively.
   Only those energy bands that are above the conduction 
   band minimum (CBM) are shown in \textbf{c} and \textbf{d}. 
 Ti-$t_{2g}$ and Nb-$t_{2g}$ states are highlighted in red. 
  The high-symmetry \textbf{k}-path of 
  BaTiO$_3$ is $\Gamma$(0,0,0)-X(0.5,0,0)-M(0.5,0.5,0)-
  $\Gamma$(0,0,0)-Z(0,0,0.5)-R(0,0.5,0.5)-A(0.5,0.5,0.5)-
  Z(0,0,0.5). The high-symmetry \textbf{k}-path 
  of LiNbO$_3$ is $\Gamma$(0,0,0)-L(0.5,0,0)-
  B(0.5,0.236,-0.236)-Z(0.5,0.5,0.5)-
  $\Gamma$(0,0,0)-X(0.368,0,-0.368).
  }
\label{fig:primitive-bs-cbm}
\end{figure}

Fig.~\ref{fig:primitive-bs-cbm} shows 
densities of states and electronic band structure
of pristine BaTiO$_3$ (panels \textbf{a} and \textbf{c}) 
and pristine LiNbO$_3$ (panels \textbf{b} and \textbf{d}) 
under a representative hydrostatic pressure of 8 GPa.
Both pristine BaTiO$_3$ and LiNbO$_3$ are wide gap
insulators under ambient pressure and they remain insulating under a
hydrostatic pressure of 8 GPa.
In the calculations, we find that the overall electronic structures
of BaTiO$_3$ and LiNbO$_3$ do not change substantially with pressure
but the band width of Ti-$d$ and Nb-$d$ states increase under
pressure. Since we are interested in the
band width of Ti-$d$ and Nb-$d$ states, we only show the electronic
band structure above the conduction band minimum (CBM). Due to the
crystal field splitting in BaTiO$_3$ and LiNbO$_3$, both Ti-$d$ and
Nb-$d$ orbitals are split into $t_{2g}$ and $e_g$
states. The Ti-$t_{2g}$ and Nb-$t_{2g}$ states are 
highlighted in red in panels \textbf{c} and \textbf{d} of
Fig.~\ref{fig:primitive-bs-cbm}. Our calculations find that the
band width of both $t_{2g}$ and $e_g$ states increase under
pressure. For conciseness, we only
show the pressure dependence of Ti-$t_{2g}$ and Nb-$t_{2g}$ band width
in Fig.~\ref{fig:ferro-rigidband} in the main text.

\newpage 
\clearpage

\section{\label{sec:polar}Definition of polar displacements}

The advantage of using polar displacements to characterize
ferroelectric-like distortions is that they are well
defined in both insulating and conducting systems, while
polarization is ill-defined in metals~\cite{PhysRevB.83.235112,
PhysRevLett.109.247601,kim2016polar}.

Pristine BaTiO$_3$ and LiNbO$_3$ are both insulators. However,
oxygen vacancies donate itinerant electrons and make oxygen-deficient
BaTiO$_{3-\delta}$ and LiNbO$_{3-\delta}$ conducting under ambient
pressure. We follow the previous definition of polar displacements when we
study BaTiO$_3$ and LiNbO$_3$ under ambient
pressure~\cite{PhysRevMaterials.3.054405} and 
now extend the definition to the
case in which both materials are under
hydrostatic pressure. For self-containedness, we outline the
definition of polar displacement $\delta z$ briefly below.

\begin{figure}[t]
\includegraphics[width=0.7\textwidth]{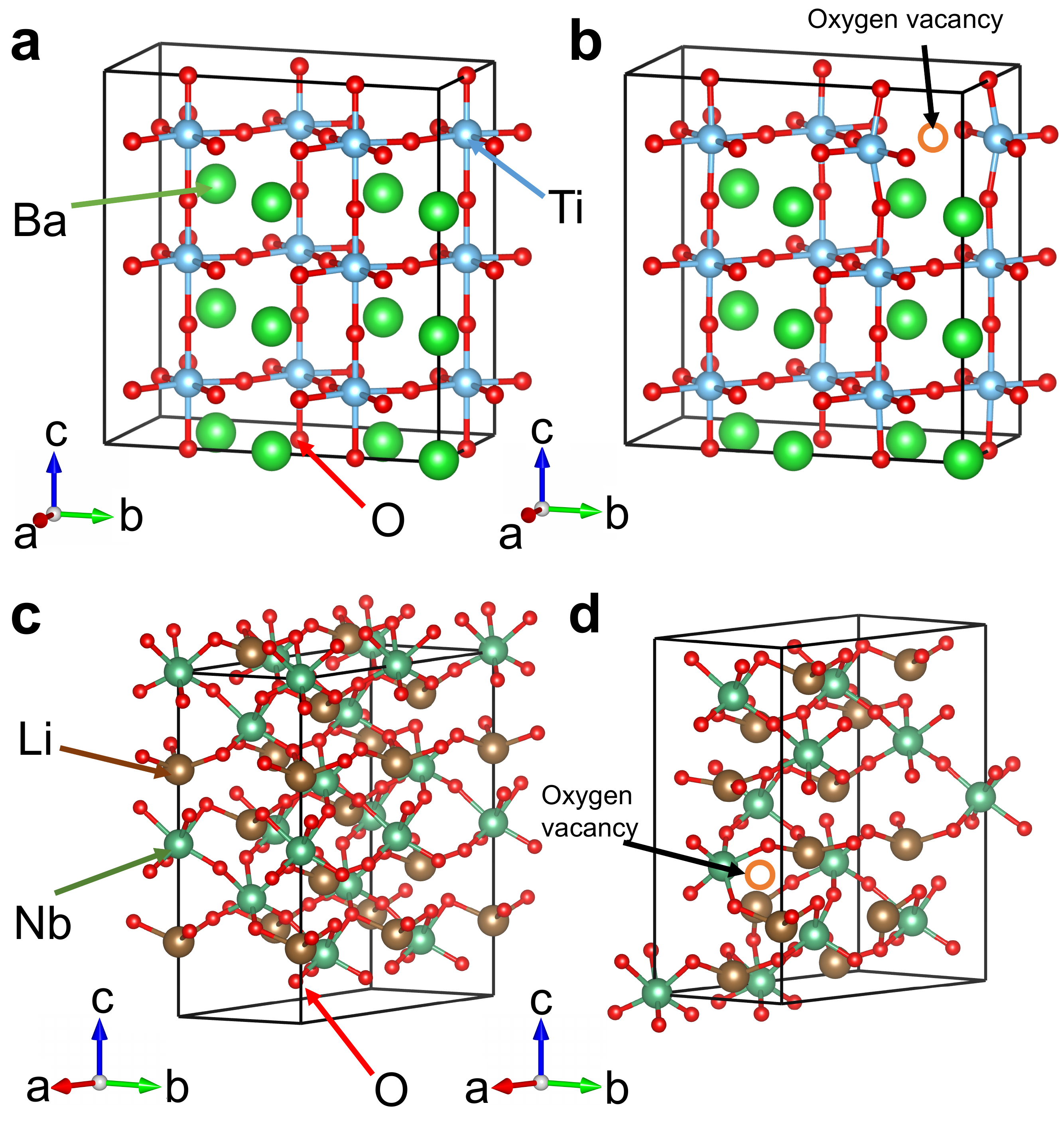}
\caption{Crystal structures of pristine BaTiO$_{3}$ (panel
  \textbf{a}), oxygen-deficient BaTiO$_{3-\delta}$ with $\delta =
  8.3\%$ (panel \textbf{b}), pristine LiNbO$_{3}$ (panel \textbf{c})
  and oxygen-deficient LiNbO$_{3-\delta}$ with $\delta = 8.3\%$ (panel
  \textbf{d}).  Green, blue and red balls in panels \textbf{a} and
  \textbf{b} are Ba, Ti and O atoms, respectively.  Brown, green and
  red balls in panels \textbf{c} and \textbf{d} are Li, Nb and O
  atoms, respectively.  The oxygen vacancy in oxygen-deficient
  BaTiO$_{3-\delta}$ and LiNbO$_{3-\delta}$ is highlighted by the
  orange open circle.}
\label{fig:LNO-BTO-structure}
\end{figure}

Fig.~\ref{fig:LNO-BTO-structure}\textbf{a} and \textbf{b} show the
crystal structure of pristine BaTiO$_{3}$ and oxygen-deficient
BaTiO$_{3-\delta}$ with $\delta = 8.3\%$.  In the $ab$ plane, each Ti
atom is surrounded by four O atoms in TiO$_2$ layer and Ti atoms move
along the $c$ axis, therefore Ti-O displacement $\delta
z_{\textrm{Ti-O}}$ is defined as:
\begin{equation}
\delta z_{\textrm{Ti-O}}= z_{\textrm{Ti}} -
\frac{1}{4}\sum_{i=1}^4z_{\textrm{O}_i}
\end{equation}     
where $z_{\textrm{Ti}}$ is the $c$ position of Ti and
$z_{\textrm{O}_i}$ is the $c$ position of the four nearest O
atoms. $\delta z_{\textrm{Ba-O}}$ is defined in a similar way as
$\delta z_{\textrm{Ti-O}}$. We note that in
oxygen-deficient BaTiO$_{3-\delta}$, the two Ti atoms that are closest
to the oxygen vacancy only have three nearest O atoms in TiO$_2$ layer
in $ab$ plane, therefore the Ti-O displacement
$\delta_{\textrm{Ti-O}}$ for these two Ti atoms is defined as:
\begin{equation}
\delta z_{\textrm{Ti-O}}= z_{\textrm{Ti}} -
\frac{1}{3}\sum_{i=1}^3z_{\textrm{O}_i}
\end{equation}     
where $z_{\textrm{O}_i}$ is the $c$ position of the three nearest O atoms
around the Ti atom in $ab$ plane.

Fig.~\ref{fig:LNO-BTO-structure}\textbf{c} and \textbf{d} show the
crystal structure of pristine LiNbO$_{3}$ and oxygen-deficient
LiNbO$_{3-\delta}$ with $\delta = 8.3\%$.  $\delta z_{\textrm{Li-O}}$
and $\delta z_{\textrm{Nb-O}}$ are defined in slightly different
ways. Each Li atom is surrounded by three O atoms and Li atoms move
along the $c$ axis.  Li-O displacement $\delta z_{\textrm{Li-O}}$ is
defined as:
\begin{equation}
\delta z_{\textrm{Li-O}}= z_{\textrm{Li}} - \frac{1}{3}\sum_{i=1}^3
z_{\textrm{O}_i} 
\end{equation}   
where $z_{\textrm{Li}}$ is the $c$ position of Li and
$z_{\textrm{O}_i}$ is the $c$ position of the three nearest O
atoms. Each Nb atom is surrounded by six O atoms and Nb atoms move
along the $c$ axis. Nb-O displacement $\delta z_{\textrm{Nb-O}}$
is defined as:
\begin{equation}
  \delta z_{\textrm{Nb-O}}= z_{\textrm{Nb}} - \frac{1}{6}\sum_{i=1}^6
  z_{\textrm{O}_i}  
\end{equation} 
where $z_{\textrm{Nb}}$ is the $c$ position of Nb and
$z_{\textrm{O}_i}$ is the $c$ position of the six nearest O
atoms. We note that in oxygen-deficient LiNbO$_{3-\delta}$, the
Li atom that is closest to the oxygen vacancy only has two nearest O
atoms, and the two Nb atoms that are closest to the oxygen vacancy
only have five nearest O atoms. Therefore, the Li-O displacement for this
Li atom and the Nb-O displacement for these two Nb atoms are defined as:
\begin{equation}
\delta z_{\textrm{Li-O}}= z_{\textrm{Li}} - 
\frac{1}{2}\sum_{i=1}^2 z_{\textrm{O}_i} 
\end{equation}   
where $z_{\textrm{O}_i}$ is $c$ position of the two nearest O atoms for the
Li atom and 
\begin{equation}
  \delta z_{\textrm{Nb-O}}= z_{\textrm{Nb}} - 
  \frac{1}{5}\sum_{i=1}^5 z_{\textrm{O}_i}  
\end{equation} 
where $z_{\textrm{O}_i}$ is $c$ position of the five nearest O atoms
around the Nb atom.

\newpage
\clearpage

\bibliographystyle{apsrev}
%\bibliography{pressure}

\end{document}